\documentclass[10pt]{iopart}
\usepackage{amssymb}
\usepackage{graphicx}
\usepackage{bm}
\usepackage{relsize}
\usepackage{subfig}
\usepackage{upgreek}
\newcommand*\nobreakhyphen{\hbox{-}\nobreak\hskip\z@skip}

\begin{document}
\title{Noise Scaling in SQUID Arrays}

\author{O. A. Nieves, K-H. M\"{u}ller}
\address{CSIRO Manufacturing, PO Box 218, Lindfield, NSW 2070, Australia}
\ead{oscar.nieves@csiro.au}
\vspace{10pt}

\begin{abstract}
We numerically investigate the noise scaling in high-$T_c$ commensurate 1D and 2D SQUID arrays. We show that the voltage noise spectral density in 1D arrays violates the scaling rule of $\sim 1/N_p$ for the number $N_p$ of Josephson junctions in parallel. In contrast, in 2D arrays with $N_s$ 1D arrays in series, the voltage noise spectral density follows more closely the expected scaling behaviour of $\sim N_s/N_p$. Additionally, we reveal how the flux and magnetic field rms noise spectral densities deviate from their expected $\sim (N_sN_p)^{-1/2}$ scaling and discuss their implications for designing low noise magnetometers.

%
\end{abstract}

%
\vspace{2pc}
\noindent{\it Keywords}: noise, SQUID, array, scaling law\newline\newline
%
\submitto{\SUST}
%
\maketitle
%
\ioptwocol

\section{\label{sec:level1}Introduction}
Superconducting quantum interference devices, better known as SQUIDs; are used extensively in magnetic sensing applications \cite{CLA06, FAG06}. When multiple SQUIDs are combined in parallel and in series to form a so-called SQUID array, the response to externally applied magnetic fields can be enhanced and tuned via the number of Josephson junctions and the geometry of the SQUID cells [3--12]. Henceforth, SQUID and superconducting quantum interference filter (SQIF) arrays can be used as highly sensitive magnetometers and low-noise amplifiers in a variety of applications [13--18].

A commonly used characteristic of operation is the voltage-flux response. The SQUID array is current biased and any small applied magnetic flux $\delta \phi$ per loop is converted into voltage $\delta \bar{v}$ across the array. The conversion efficiency is given by the transfer function $\bar{v}_{\phi} = \partial \bar{v} / \partial \phi$.
The electrical normal resistance of Josephson junctions (JJs) generates Johnson white noise, which causes the appearance of voltage and flux noise in SQUID arrays. Both the transfer function and the noise spectral densities depend on many device parameters. The problem of optimising the common dc SQUID has long been solved [19--21]. In contrast, optimising SQUID arrays is still a partially unsolved problem due to the larger parameter space and computational complexity. The transfer function of 1D and 2D arrays has been studied theoretically \cite{GAL22}, but their noise spectral densities have not been simulated yet.
 
The current paper is organised as follows. In Sec. II we investigate the maximum transfer function of commensurate SQUID arrays with $N_p = 2  \textendash   20$ JJs in parallel and $N_s = 1   \textendash    20$ JJ rows in series for the case where temperature, critical current, normal resistance and partial inductances are kept constant. In Sec. III we explore the low-frequency voltage noise spectral density and the rms flux and magnetic field noise spectral densities, and their deviation from the expected scaling. Finally, we discuss some of the implications this has for the design of high-$T_c$ SQUID arrays.

\section{\label{sec:level1}Array transfer function}
We start by discussing the transfer function of 1D and 2D SQUID arrays, since the transfer function is needed to calculate the flux noise and magnetic field noise spectral densities. We assume that the JJs of the SQUID arrays are over-damped, a valid assumption for YBCO thin film arrays at 77 K, and all arrays have the same normal resistances $R$ and critical currents $I_c$, that is: there is no statistical variation in the junction parameters. The time-averaged voltage, appearing between the top and bottom bias current leads (Fig. 1), is $\bar{v}$ and is normalised by $R I_c$. The transfer function $\bar{v}_\phi$ of a SQUID array is $\bar{v}_{\phi} = \partial \bar{v} / \partial \phi_a$, where $\phi_a = \Phi_a/\Phi_0$ with $\Phi_a$ the applied flux per SQUID cell and $\Phi_0$ the flux quantum. The transfer function $\bar{v}_\phi$ depends on several parameters, which can be grouped into intrinsic, extrinsic and geometric parameters, where
\begin{equation}\label{eq:transfer-function}
\bar{v}_{\phi} = \bar{v}_{\phi}(I_c, T, I_b, \Phi_a, \hat{L}, N_s, N_p)   .
\end{equation}
Here, $I_c$ is the only intrinsic parameter as $R$ has been absorbed by normalisation. The three external parameters are the applied temperature $T$, the applied total bias current $I_b$ and the flux $\Phi_a$ applied per SQUID cell. The geometrical parameters are the inductance matrix $\hat{L}$ of the commensurate array, the number $N_s$ of JJ rows in series and the number $N_p$ of JJs in parallel in each row. We fix $T$ at 77 K which is common for YBCO devices.

To obtain optimal flux-to-voltage transduction, the transfer function of the SQUID array can be maximised by adjusting the external bias current $I_b$ and the external applied flux $\Phi_a$ such that the transfer function is at its maximum. We denote the maximum transfer function at $i_b=i_b^*$ and $\phi_a=\phi_a^*$ by $\bar{v}_{\phi}^{\mathrm{m}\mathrm{a}\mathrm{x}}$. 

\begin{figure}[h]\begin{center}
\includegraphics[width=0.45\textwidth]{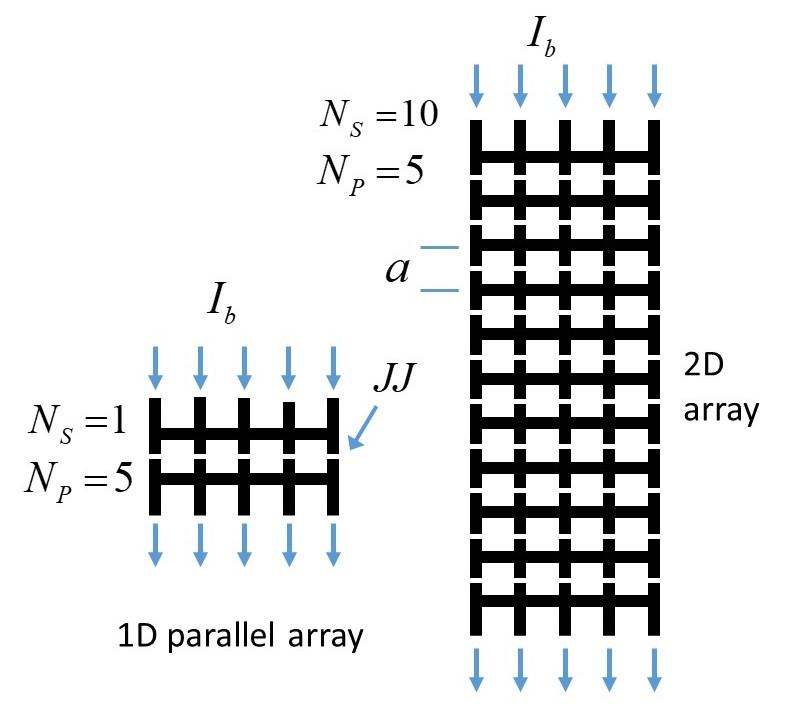}
\caption{Examples of commensurate thin-film SQUID arrays. On the left, the layout of an $N_p = 5, N_s = 1$, 1D SQUID array and on the right a 2D SQUID array with $N_p = 5, N_s = 10$. The SQUID loop areas are squares with side length $a$, which is the cell size. The location of the JJs is indicated as gaps. The total
biasing current $I_b$ is injected uniformly from the top with each of the $N_p$ leads receiving the same constant current $I_b / N_p$. A spatially homogeneous magnetic flux $\phi_a$ per SQUID cell is applied to the array.} 
\label{fig:diagram}
\end{center}
\end{figure}

\begin{figure}[h]\begin{center}
\includegraphics[width=0.45\textwidth]{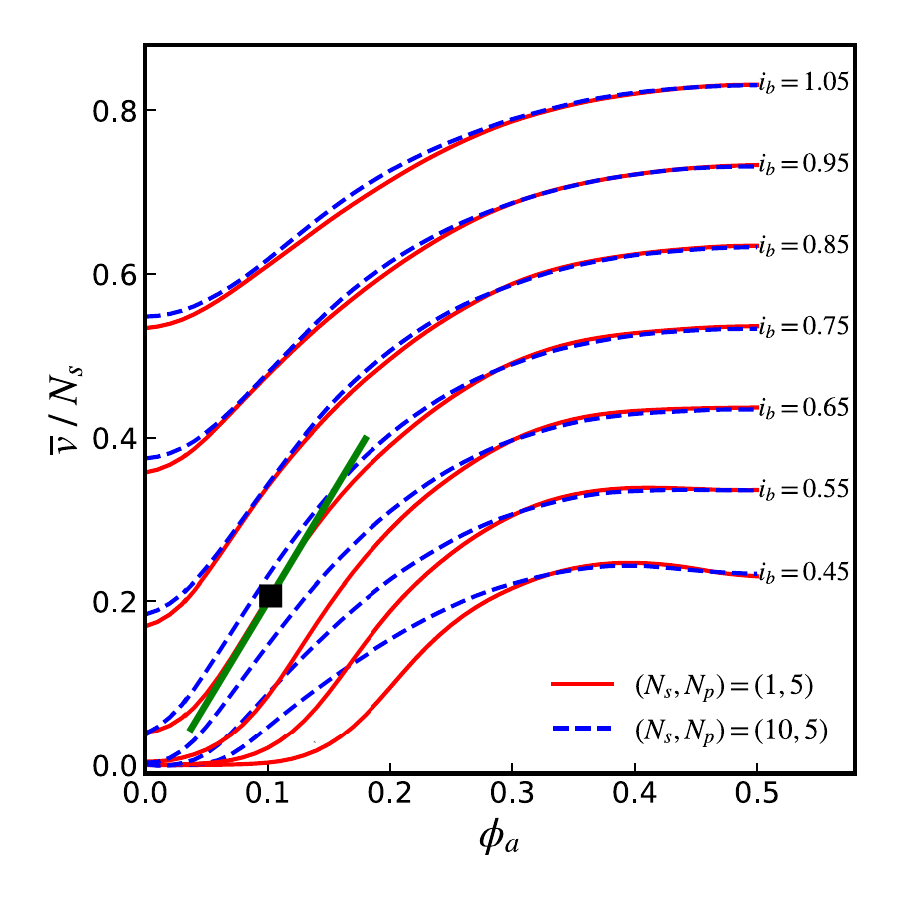}
\caption{Normalised voltage-flux response of a SQUID array with $N_p = 5$, for different normalised bias currents $i_b$ where $N_s = 1$ (1D parallel array, red) and $N_s = 10$ (2D array, dashed blue). Similar to the common dc SQUID, $\bar{v}$ is symmetric about the origin and translation invariant with period 1. In green we show the maximum slope, defining $\bar{v}_{\phi}^{\mathrm{m}\mathrm{a}\mathrm{x}}$ for (1,5). The black square denotes $\phi_a^*$ where $\bar{v}_{\phi}$ has its maximum.}
\label{fig:VB-response}
\end{center}
\end{figure}

\begin{figure}[!h]\begin{center}
\includegraphics[width=0.45\textwidth]{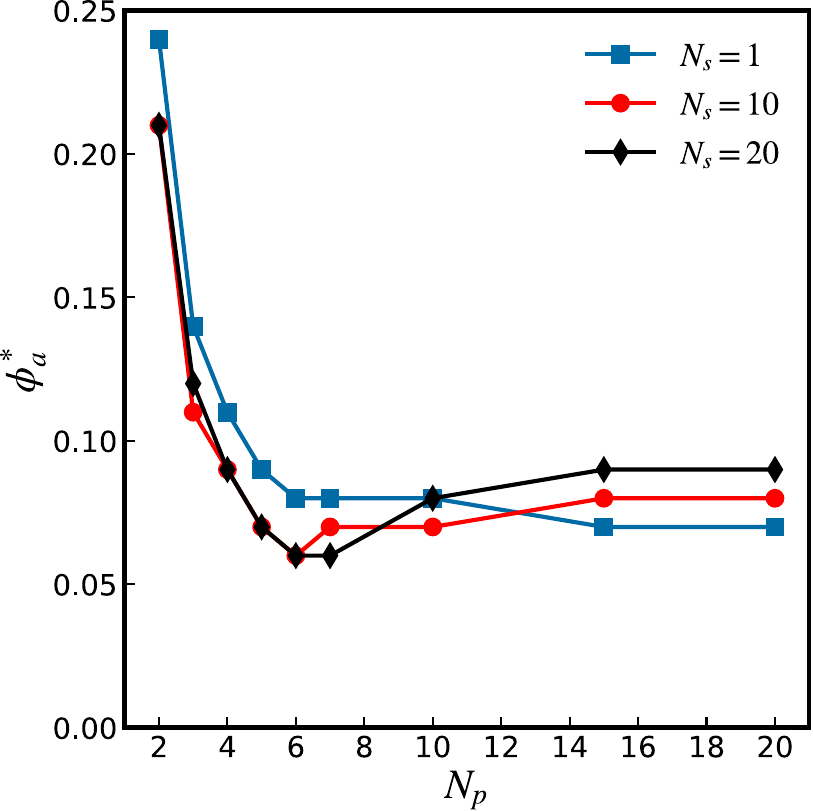}
\caption{Normalised applied flux $\phi^*_a$ per SQUID loop which maximises the transfer function $\bar{v}_\phi$ versus the number $N_p$ of
JJs in parallel, for $N_s = 1$ (1D parallel arrays) and $N_s = 10$ and 20 (2D arrays).}
\label{fig:optimal-flux}
\end{center}
\end{figure}
\begin{figure}[!h]\begin{center}
\includegraphics[width=0.45\textwidth]{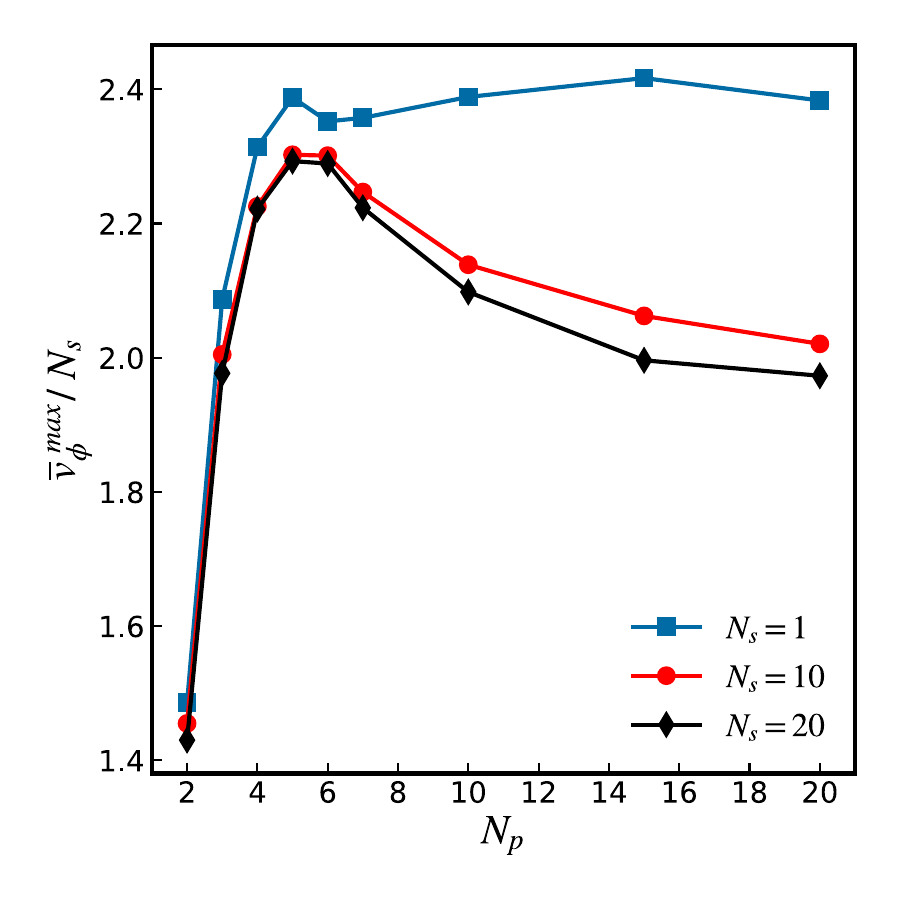}
\caption{Maximum transfer function $\bar{v}_{\phi}^{\mathrm{m}\mathrm{a}\mathrm{x}} / N_s$ versus $N_p$ for $N_s = 1$ (1D parallel arrays) and $N_s = 10$ and $20$ (2D arrays).}
\label{fig:transfer-function}
\end{center}
\end{figure}

The theoretical model used here to calculate the maximum transfer functions $\bar{v}_{\phi}^{\mathrm{m}\mathrm{a}\mathrm{x}}$ and the spectral noise densities is simlar to the RSJ simulation model used by Cybart {\it{et al.}} \cite{CYB12}. The mathematical model used here has been discussed in detail in a separate publication \cite{GAL22}, which takes into account the effect of thermal noise and mutual inductances.

In the following simulations, we keep $I_c$ and $\hat{L}$ fixed while varying the array geometric parameters $N_s$ and $N_p$. We use $I_c$ = 20 $\upmu$A, which is a typical value for YBCO step edge JJs \cite{MIT19}. The inductance matrix $\hat{L}$ is defined by the commensurate array layout shown in Fig. \ref{fig:diagram}. Here, the square loop width is $a$ = 10 $\upmu$m with track width $w$ = 2 $\upmu$m and film thickness 0.2 $\upmu$m. The inductance matrix includes the kinetic inductances where the London penetration depth is taken as $\lambda = 0.4    \upmu$m. For the geometric part of the inductance matrix, we use the analytic expressions given in \cite{HOE65}. From the self-inductance $L_s$ of the individual SQUID loops, one finds $\beta_L = 2 I_c L_s / \Phi_0 = 0.76$. The noise strength parameter corresponding to these values and $T=77$ K is $\Gamma = 2\pi k_B T/(\Phi_0 I_c) = 0.16$, where $k_B$ is the Boltzmann constant. The top and bottom bias leads were taken as 100 $\upmu$m long and their inductances were included in our calculations, though their contributions were found to be negligibly small. 

As an example, Fig. \ref{fig:VB-response} shows for $N_p = 5$ the time-averaged voltage $\bar{v} / N_s$, for $N_s = 1$ (red) and $N_s = 10$ (dashed blue), versus the applied flux $\phi_a$ for different bias currents $i_b = I_b / (N_p I_c)$ where $I_b$ is the total bias current (see Fig. \ref{fig:diagram}). We see that $\bar{v} / N_s$ depends on $i_b$. In the case of $N_s = 1$, the maximum transfer function $\bar{v}_{\phi}^{\mathrm{m}\mathrm{a}\mathrm{x}}$ occurs at $i_b^* =0.75$ and $\phi^*_a = 0.095$.

By varying $N_s$ and $N_p$, we find that $\bar{v}_{\phi}^{  max}$ occurs at $i_b^* \approx 0.75$, independent of $N_s$ and $N_p$. In contrast, the applied flux $\phi_a^*$ varies strongly with $N_p$. As shown in Fig. \ref{fig:optimal-flux}, $\phi_a^*$ initially rapidly decreases with increasing $N_p$. While $\phi_a^*$ = 0.25 for the common dc SQUID ($N_s = 1, N_p = 2$), $\phi_a^*  \approx 0.075$ if $N_p \gtrsim  6$ for both $N_s =1$ (1D parallel arrays) as well as $N_s = 10$ and 20 (2D arrays).

Figure \ref{fig:transfer-function} shows $\bar{v}_{\phi}^{  max} / N_s$ versus $N_p$ for $N_s= 1$ (1D parallel arrays) and $N_s = 10$ and 20 (2D arrays).  The maximum transfer function $\bar{v}_{\phi}^{ max}$ initially increases with $N_p$. However, for $N_p \gtrsim 6$, $\bar{v}_{\phi}^{\mathrm{m}\mathrm{a}\mathrm{x}}$ plateaus for the 1D parallel arrays and slightly decreases for the 2D arrays. The levelling of $\bar{v}_{\phi}^{  max} / N_s$ for $N_p \gtrsim 6$ in Fig. \ref{fig:transfer-function} can be understood from calculations performed by Kornev {\it{et al.}} \cite{KOR09b, KOR11} and others \cite{GAL21}.

\section{\label{sec:level1}Array low-frequency voltage and flux noise spectral density}
The voltage noise spectral density can be obtained from the Fourier transform of $v(\tau)$. The one-sided voltage noise spectral density $S_v(f)$ in dimensionless units (normalised by $R I_c \Phi_0 / (2 \pi)$) is given by 
\begin{equation}\label{eq:Sv}
S_v(f) = \lim_{\tau_0   \to   \infty} \frac{2}{\tau_0}  \left|\int_{- \tau_0 / 2}^{\tau_0 /2} v(\tau)   e^{i   2 \pi f  \tau} d\tau \right|^2,
\end{equation}
where $i$ is the unit imaginary number. Here, $v(\tau)$ is the time-dependent total voltage between the bias leads of the array, $f$ is the spectral frequency in dimensionless units, normalised by $2 \pi R I_c / \Phi_0$, and $\tau$ the time in dimensionless units, normalised by $\Phi_0 / (2 \pi R I_c)$.

The low-frequency voltage-noise spectral density $S_v(0)$ can be calculated from a low frequency analysis as detailed in Tesche \& Clarke \cite{TES77}. This can be done by averaging the total array voltage $v(\tau)$ of the array, i.e.
\begin{equation}\label{eq:vtau}
v(\tau)= \sum_{j=1}^{N_s} \frac{1}{N_p}    \sum_{k=1}^{N_p} v_{j k}(\tau)   ,
\end{equation}
over time intervals $\Delta \tilde{\tau} = 400 \Delta \tau$ and obtaining a set of $N = 512$ averaged voltages \cite{ENP85a}, where $v_{j k}(\tau)$ is the voltage corresponding to the $k$th junction (from left to right) in the $j$th row of the array. We then take the Fourier transform of this discrete set and by using Eq.  \ref{eq:Sv} the low-frequency voltage-noise $S_v(0)$ is determined. To enhance the numerical accuracy, we repeat this process up to 7000 times in order to achieve a good ensemble average for $S_v(0)$. This discrete Fourier transform procedure is accurate subject to the condition $f_J \ll 1/ \Delta \tilde{\tau} \ll N f_J$ \cite{TES77}, where $f_J$ is the normalised fundamental Josephson frequency $\bar{v}(i_b^*,\phi_a^*) / 2\pi$. In our simulations, we use $\Delta \tau = 0.01$. In this paper, we limit the size of our arrays to no more than $(20,20)$ due to the large computation times required to accurately compute $S_v$.

As the Johnson noise voltages of the JJ's are uncorrelated and their mean square deviations are identical, one would simplistically expect to obtain the scaling behaviour
\begin{equation}\label{eq:Sv_scaling}
S_v(0)   \propto   \frac{N_s}{N_p} \; .
\end{equation}
This is evident from Eq.  \ref{eq:Sv} and \ref{eq:vtau}: $v(\tau)$ is calculated by taking the arithmetic mean of the junction voltages in parallel, and then summing up the voltage in series for $N_s$ rows.

Similar to $S_v(0)$, the voltage to voltage-noise ratio SNR$_{ v}$ is expected to follow the scaling behaviour
\begin{equation}\label{eq:SNR_v_scaling}
SNR_{ v}    \propto    \frac{N_s}{(N_s / N_p)^{1/2}} = (N_s N_p)^{1/2} \; .
\end{equation}

Using our above mentioned 2D SQUID array model, we have calculated the normalised low-frequency voltage-noise spectral density $S_v(0)$ from Eq.\ref{eq:Sv} for 1D parallel arrays and 2D arrays. 

\begin{figure}[!h]\begin{center}
\includegraphics[width=0.45\textwidth]{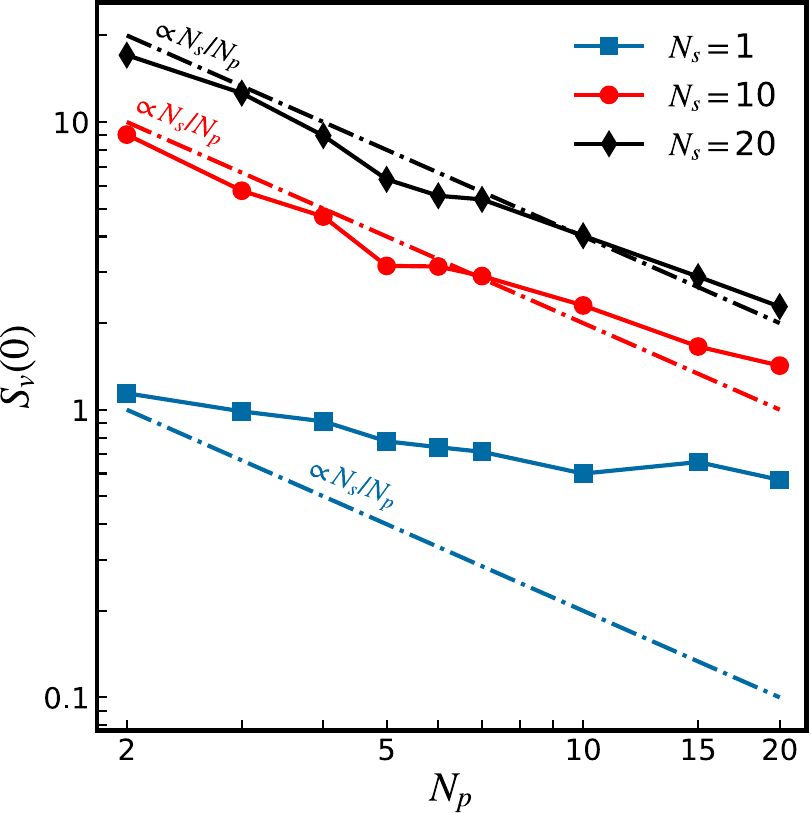}
\caption{Low-frequency voltage-noise spectral density $S_v(0)$ versus $N_p$ for $N_s = 1$ (1D parallel arrays) and $N_s =10$ and 20 (2D arrays). The dashed green curves show the scaling $N_s / N_p$ of Eq.  \ref{eq:Sv_scaling}.}
\label{fig:Sv-vs-Np}
\end{center}
\end{figure}

Figure \ref{fig:Sv-vs-Np} shows $S_v(0)$ versus $N_p$ for $N_s = 1$ (1D parallel arrays) and $N_s =10$ and 20 (2D arrays) calculated at $\phi_a^*$ (Fig. \ref{fig:optimal-flux}) and $i_b^*$ where the transfer functions $\bar{v}_{\phi}$ have their maxima.
The dashed curves indicate the $N_s / N_p$ scaling behaviour. The calculation shows that the voltage noise spectral density $S_v(0)$ for the 1D parallel arrays does not follow the $N_s / N_p$ scaling but instead, $\sim N_s/N_p^{0.3}$. In contrast, the 2D arrays follow the $N_s / N_p$ scaling fairly well.

\begin{figure}[!h]\begin{center}
\includegraphics[width=0.45\textwidth]{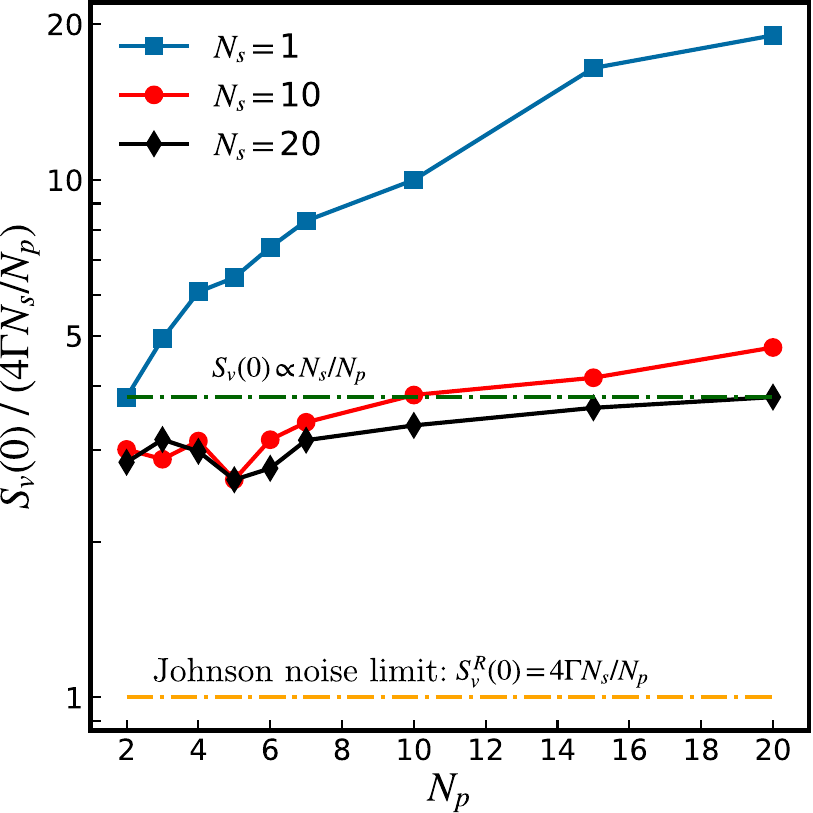}
\caption{Low-frequency normalised voltage noise spectral density $S_{v}(0) / (4   \Gamma N_s / N_p)$ versus $N_p$ for $N_s = 1$ (1D parallel arrays) and $N_s =10$ and 20 (2D arrays). The dashed green curve shows the scaling $N_s / N_p$ of Eq. (5). In orange the Johnson noise limit.}
\label{fig:Sv-normalised}
\end{center}
\end{figure}

It is also useful to plot $S_v(0)$ relative to the normalised Johnson white noise voltage spectral density $S_v^R(0)$ for a purely resistive array with resistance $N_s R / N_p$. Since the de-normalised white-noise voltage spectral density of a resistor $R$ is $4 k_B T R$ \cite{NYQ28}, one finds $S_v^R(0) =  4   \Gamma   N_s /  N_p$ where $\Gamma$ is the noise strength. Using the data from Fig. \ref{fig:Sv-vs-Np}, Fig. \ref{fig:Sv-normalised} shows $S_v(0) / ( 4   \Gamma N_s / N_p) = S_v(0) / S_v^R(0)$ versus $N_p$ for different $N_s$. Figure \ref{fig:Sv-normalised} clearly reveals the deviations from the $N_s / N_p$ scaling, where for perfect scaling the data would follow horizontal lines like the dashed green line.
In particular, the 1D parallel arrays ($N_s = 1$, in red) do not follow the scaling. The orange dashed horizontal line in Fig. \ref{fig:Sv-normalised} is the Johnson noise limit, i.e.  $S_v^R(0) =  4   \Gamma   N_s /  N_p$, and the 2D SQUID arrays with relatively large $N_s$ get closest to this limit.

Kornev {\it{et al.}} \cite{KOR97, KOR09b} have shown that such a behaviour for $S_v(0)$ in 1D parallel arrays occur due to the emergence of a finite JJ interaction radius \cite{KOR11} but they did not examine the behaviour of 2D arrays.

In practice, the rms flux noise $S_{\phi}^{1/2}(f)$ is used as a measure of the device's performance. It is given by the expression
\begin{equation}\label{eq:Sphi}
S_{\phi}^{1/2}(f) = \frac{S_v^{1/2}(f)}{\bar{v}_{\phi}}   .
\end{equation}
Since $\bar{v}_{\phi}$ approximately scales with $N_s$, one expects for $S_{\phi}^{1/2}(0)$ the scaling behaviour
\begin{equation}\label{eq:Sphi_scaling}
S_{\phi}^{1/2}(0) \propto (N_s N_p)^{-1/2} \; ,
\end{equation}

\noindent
and for the flux to flux-noise ratio, SNR$_{  \phi}$, 
\begin{equation}\label{eq:SNR_phi_scaling}
 SNR_{  \phi} \propto (N_s N_p)^{1/2} \; ,
\end{equation}
\noindent
which is the same scaling as for SNR$_{  v}$ in Eq.  \ref{eq:SNR_v_scaling}. 

\begin{figure}[!h]\begin{center}
\includegraphics[width=0.45\textwidth]{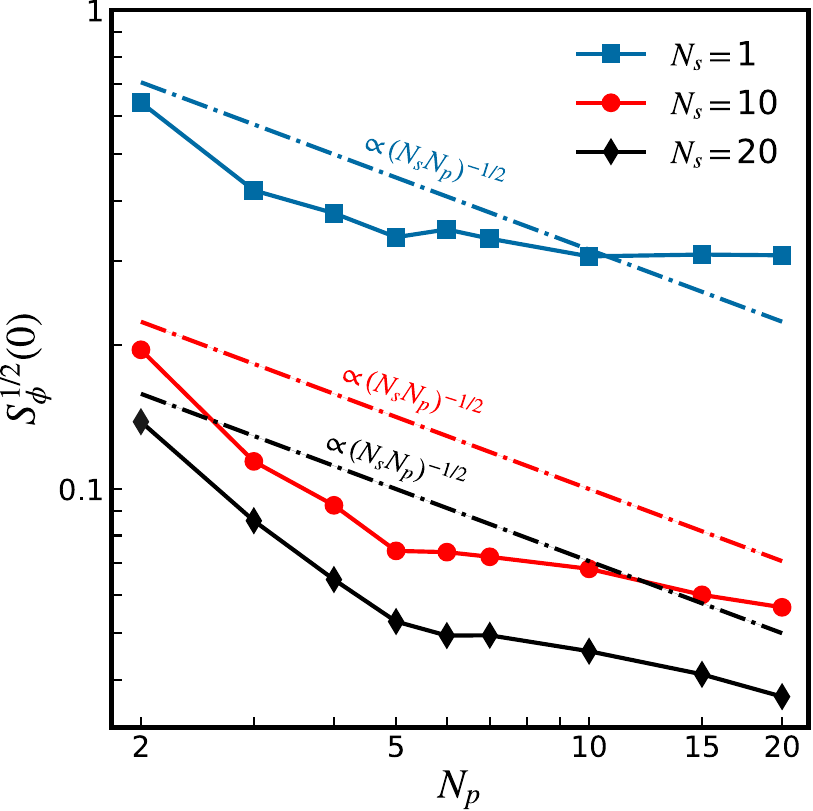}
\caption{Low-frequency rms flux noise $S_{\phi}^{1/2}(0)$ versus $N_p$ for $N_s = 1$ (1D parallel arrays) and $N_s =10$ and $20$ (2D arrays). The dashed lines show the scaling $(N_s N_p)^{-1/2}$.}
\label{fig:Sphi-vs-Np}
\end{center}
\end{figure}

Figure \ref{fig:Sphi-vs-Np} shows the calculated low-frequency rms flux noise $S^{1/2}_{\phi}(0)$ versus $N_p$ for different $N_s$. The $S^{1/2}_{\phi}$ were obtained from Eq.  \ref{eq:Sphi} at $\phi^*_a$ and $i_b^*$. As can be seen, for the three different $N_s$ the deviations from the $S^{1/2}_{\phi} \propto (N_s N_p)^{-1/2}$ scaling (dashed straight lines) are similar. This is due to the $\bar{v}_{\phi}^{\; -1}$ factor in Eq.  \ref{eq:Sphi}.

A revealing measure for the rms flux noise of a SQUID array is the dimensionless quantity $\xi_\Phi^{1/2}$ defined as
\begin{equation}\label{eq:xi}
\xi_\phi^{1/2} = \frac{S_{\phi}^{1/2}(0)}{\left( 4   \Gamma / N_s N_p \right)^{1/2} } \; .
\end{equation}
\begin{figure}[!h]\begin{center}
\includegraphics[width=0.45\textwidth]{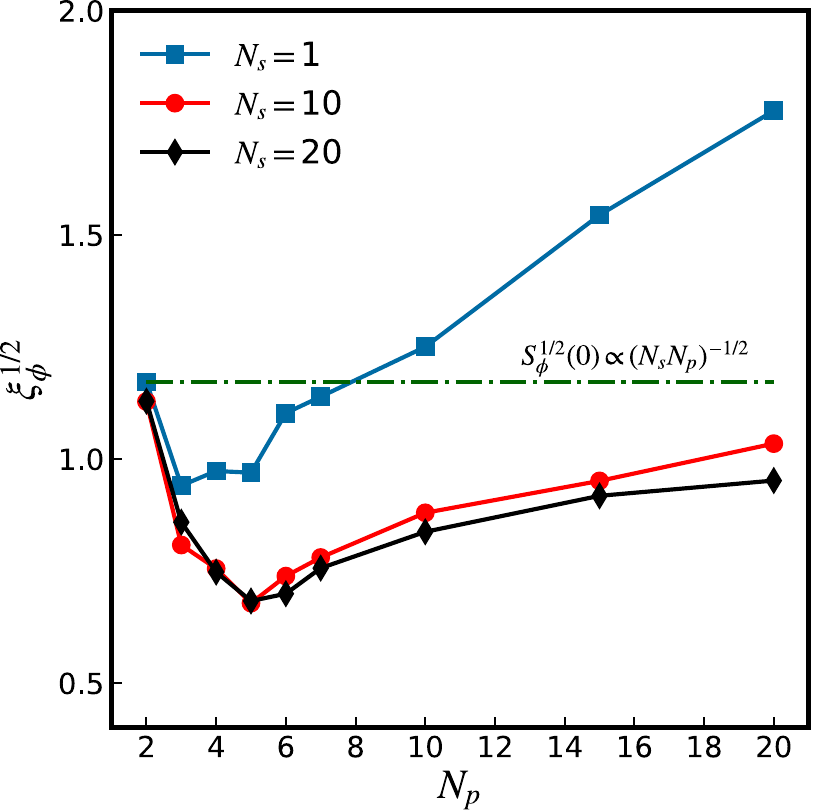}
\caption{Low-frequency rms flux noise measure $\xi_{\phi}^{1/2}$ defined by Eq. \ref{eq:xi} versus $N_p$ for $N_s = 1$ (1D parallel arrays) and $N_s =10$ and 20 (2D arrays). The dashed green curve shows the scaling $(N_s N_p)^{-1/2}$ of Eq. \ref{eq:Sphi_scaling}.}\label{default}
\label{fig:Sphi-normalised}
\end{center}
\end{figure}

The result for Eq.  \ref{eq:xi}, evaluated at $\phi^*_a$ and $i_b^*$, is displayed in Fig. \ref{fig:Sphi-normalised} showing $\xi_\phi^{1/2}$ versus $N_p$ for different $N_s$. Compared to Fig. \ref{fig:Sphi-vs-Np}, Fig. \ref{fig:Sphi-normalised} reveals the relative deviation from the $\sim (N_s N_p)^{-1/2}$ scaling. In the case of perfect scaling, the data would lie on horizontal straight lines similar to the green dashed line.

\begin{figure}[!h]\begin{center}
		\hspace*{-7mm}
		\includegraphics[width=0.45\textwidth]{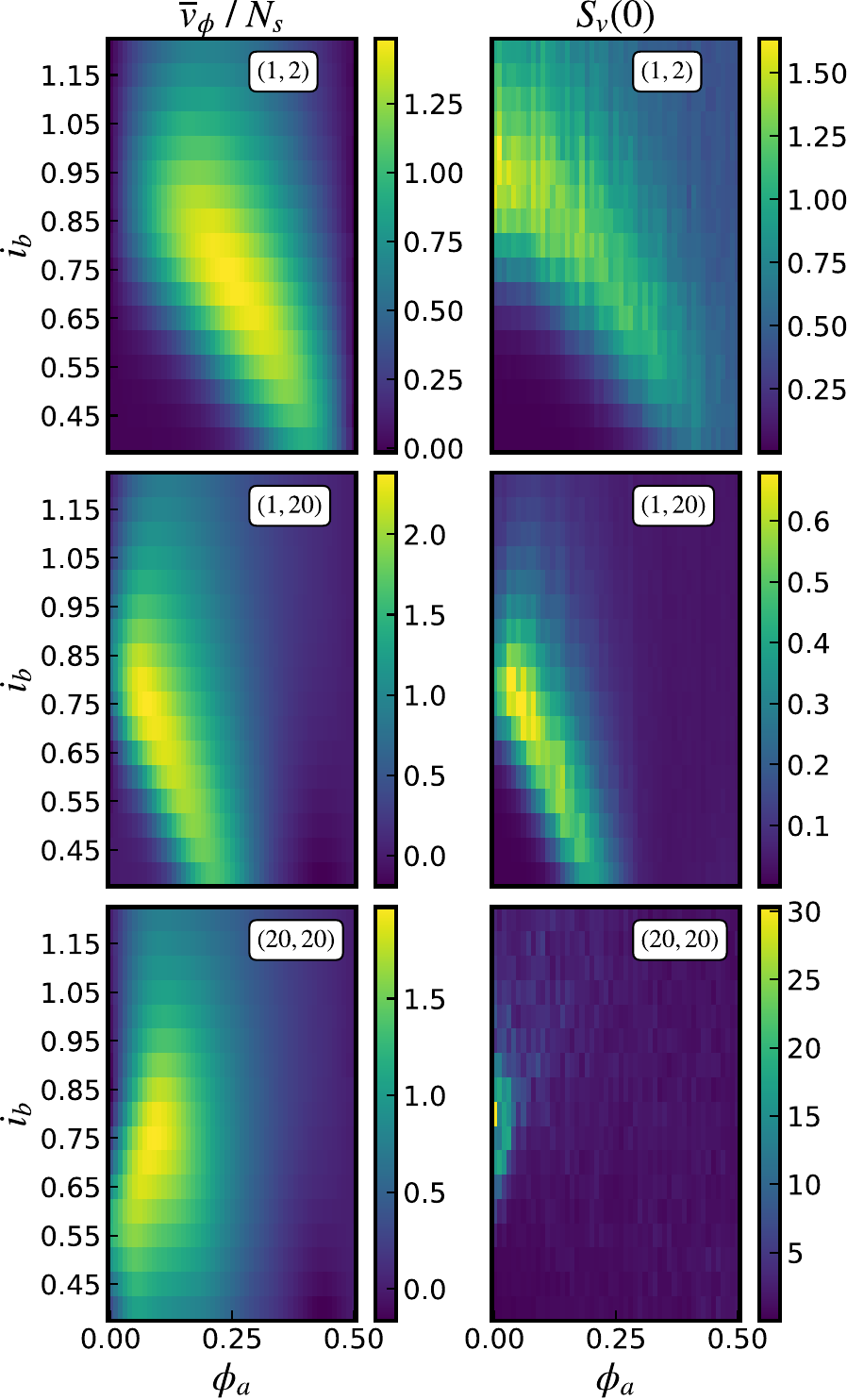}
		\caption{Normalised transfer function and voltage noise spectral density as functions of $i_b$ and $\phi_a$ for SQUID arrays of varying sizes.}
  \label{fig:Sv-heatmap}
	\end{center}
\end{figure}

\begin{figure}[!h]\begin{center}
		\hspace*{-7mm}
		\includegraphics[width=0.45\textwidth]{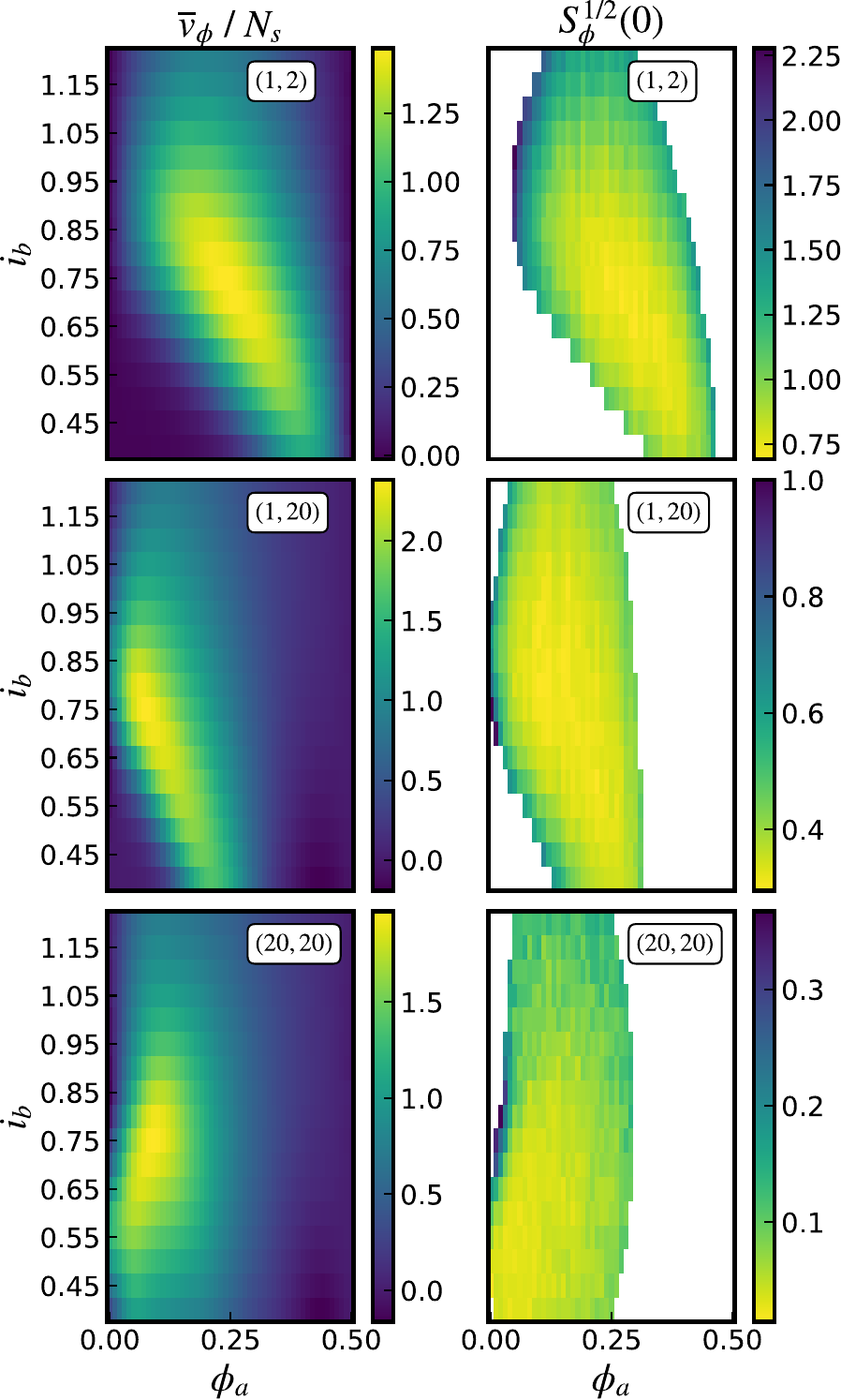}
		\caption{Normalised transfer function and flux noise spectral density as functions of $i_b$ and $\phi_a$ for SQUID arrays of varying sizes. The blank white regions correspond to points where $\bar{v}_{\phi}/N_s\rightarrow 0$ and have been excluded from the plots to improve visibility of the smaller $S_{\phi}^{1/2}(0)$ values.}
        \label{fig:Sphi-heatmap}
	\end{center}
\end{figure}

It is important to note that the values $(i_b^*, \phi_a^*)$ which maximise $\bar{v}_{\phi}$ do not minimise the voltage noise $S_v(0)$. Figure \ref{fig:Sv-heatmap} shows the distribution of $\bar{v}_{\phi}$ and $S_v(0)$ values for a range of $i_b$ and $\phi_a$, not just $i_b^*$ and $\phi_a^*$. Similarly, one can plot $S_{\phi}^{1/2}(0)$ for multiple $(i_b, \phi_a)$ and see how it compares to $\bar{v}_{\phi}$. Figure \ref{fig:Sphi-heatmap} shows several $S_{\phi}^{1/2}$ heatmaps for differently-sized arrays. They show that $S_{\phi}^{1/2}(0)$ is approximately minimised in the neighbourhood where $\bar{v}_{\phi}$ is a maximum for the (1,2) and (1,20)-arrays. However, this is not the case for the (20,20)-array, which indicates that one cannot optimise both the transfer function and noise of the array with the same $(i_b^*,\phi_a^*)$ values for arrays of arbitrary size.

\begin{figure}[!h]\begin{center}
\includegraphics[width=0.45\textwidth]{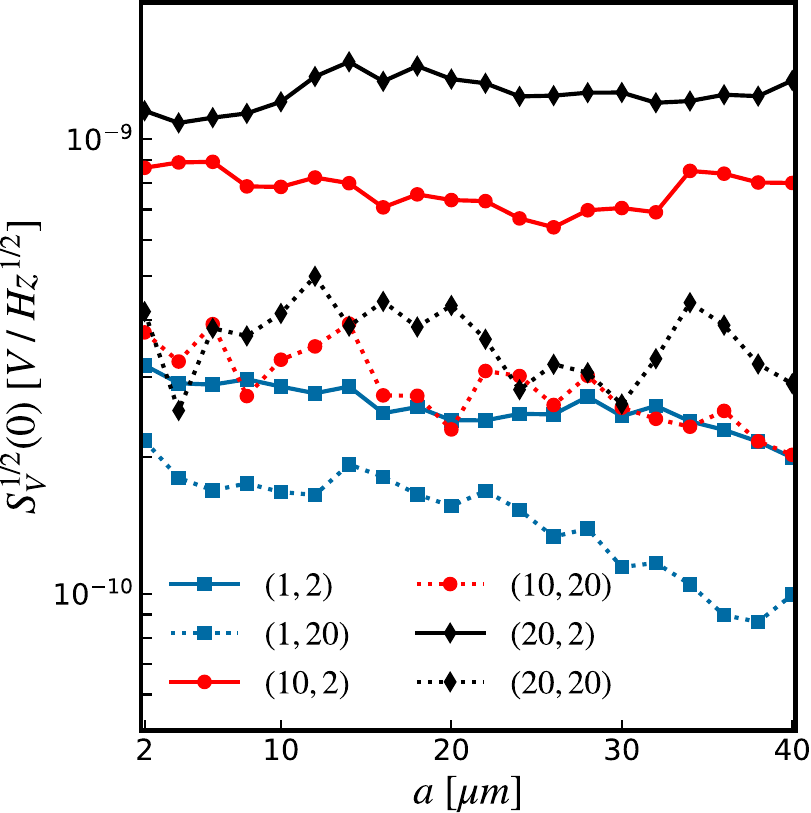}
\caption{RMS voltage noise spectral density at $T=77$ K for different cell sizes $a$ and array configurations. The range of $a$ corresponds to a $\beta_L$ range from 0.1 to 3.8.}
\label{fig:voltage-loops}
\end{center}
\end{figure}

We now proceed to de-normalise the normalised voltage noise spectral density $S_v$. This is done by multiplying $S_v$ by $RI_c\Phi_0/2\pi$. Using $R=10 \Omega$ \cite{CLA06}, we compute the rms voltage spectral density $S_V^{1/2}$ for different SQUID cell sizes $a$ and show the results in Fig. \ref{fig:voltage-loops}. 

\begin{figure}[!h]\begin{center}
\includegraphics[width=0.45\textwidth]{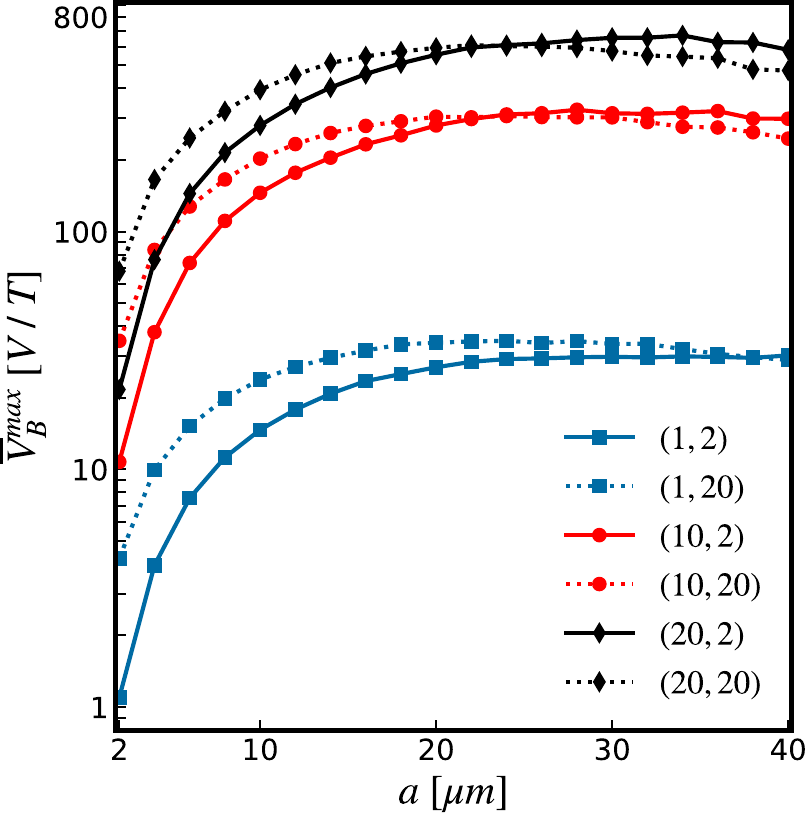}
\caption{Maximum transfer function at $T=77$ K for a range of cell sizes $a$ and array configurations. The range of $a$ corresponds to a $\beta_L$ range from 0.1 to 3.8.}
\label{fig:transfer_function-loops}
\end{center}
\end{figure}

In a similar manner, one can de-normalise the normalised transfer function $\bar{v}_{\phi}$ to obtain $\overline{V}_B=\partial\overline{V}/\partial B_a$. This is achieved by multiplying $\bar{v}_{\phi}$ by $RI_cA_{\mathrm{e}\mathrm{f}\mathrm{f}}/\Phi_0$ where $A_{\mathrm{e}\mathrm{f}\mathrm{f}}$ is the effective area of the SQUID cell, which in our case is $A_{\mathrm{e}\mathrm{f}\mathrm{f}}=a^2$. The results for $\overline{V}_B^{\mathrm{m}\mathrm{a}\mathrm{x}}$ as a function of $a$ are shown in Fig. \ref{fig:transfer_function-loops}.

The normalised rms flux noise $S_{\phi}^{1/2}$ is de-normalised by multiplying $S_{\phi}^{1/2}$ with $\gamma = \Phi_0^{3/2}/\sqrt{2 \pi R I_c}$. Using again the junction parameters at $T=77$ K of $R=10 \Omega$ and $I_c=20 \upmu$A one obtains $\gamma = 1.28   \upmu \Phi_0 / \sqrt{\mathrm{H}\mathrm{z}}$. Using the normalised $S_{\phi}^{1/2}$ value from Fig. \ref{fig:Sphi-heatmap} corresponding to the $(20,20)-$array, one finds an rms flux noise of $0.05 \upmu \Phi_0 / \sqrt{\mathrm{H}\mathrm{z}}$. In contrast, the rms flux noise of the $(1,20)$-array is 10 times higher, while according to the scaling behaviour in Eq. (8) it should be $\approx 4.5$ times higher.

\begin{figure}[h]\begin{center}
\includegraphics[width=0.45\textwidth]{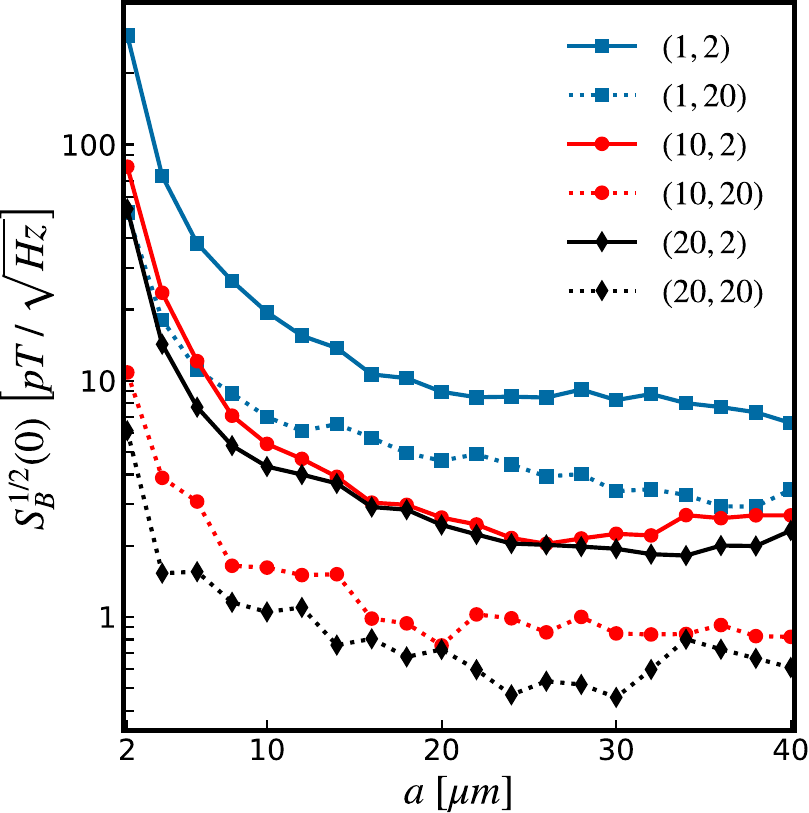}
\caption{Magnetic field noise spectral density at $T=77$ K for a range of cell sizes $a$ and array configurations. The range of $a$ corresponds to a $\beta_L$ range from 0.1 to 3.8.}
\label{fig:SB-loops}
\end{center}
\end{figure}

The rms magnetic field noise spectral density $S_B^{1/2}$ is obtained by dividing the rms flux noise by the SQUID loop area $a^2$, or simply by calculating $S_V^{1/2}/\overline{V}_B^{\mathrm{m}\mathrm{a}\mathrm{x}}$. This gives for the $N_s = N_p =20$ array with $a^2=100  \upmu\mathrm{m}^2$ a value of $S_B^{1/2} = 1.0   \mathrm{p}\mathrm{T} / \sqrt{\mathrm{H}\mathrm{z}}$. This result is consistent with the literature, for instance Cou\"{e}do {\textit{et al.}} \cite{COU19} reported a white noise measurement of $\approx 300$ fT$/\sqrt{\mathrm{H}\mathrm{z}}$ on a (300,2)-SQIF array made of YBCO and operating at $T=66$ K. By contrast, in low-temperature dc-SQUIDs operating at $T\leq 4.2$K, one often finds $S_B^{1/2}\approx 1-4 \mathrm{f}\mathrm{T} / \sqrt{\mathrm{H}\mathrm{z}}$ using pick-up coils (see for instance Drung \textit{et al.} \cite{DRU91, DRU07}). Similarly, earlier work on high-$T_c$ YBCO dc-SQUIDs operating at 77 K showed that by coupling the SQUID to a large pickup loop of millimetre size, that $S_B^{1/2}$ could be reduced down to $\approx 10$ fT$/\sqrt{\mathrm{H}\mathrm{z}}$ \cite{KOE99}.
\begin{figure}[!h]\begin{center}
\includegraphics[width=0.45\textwidth]{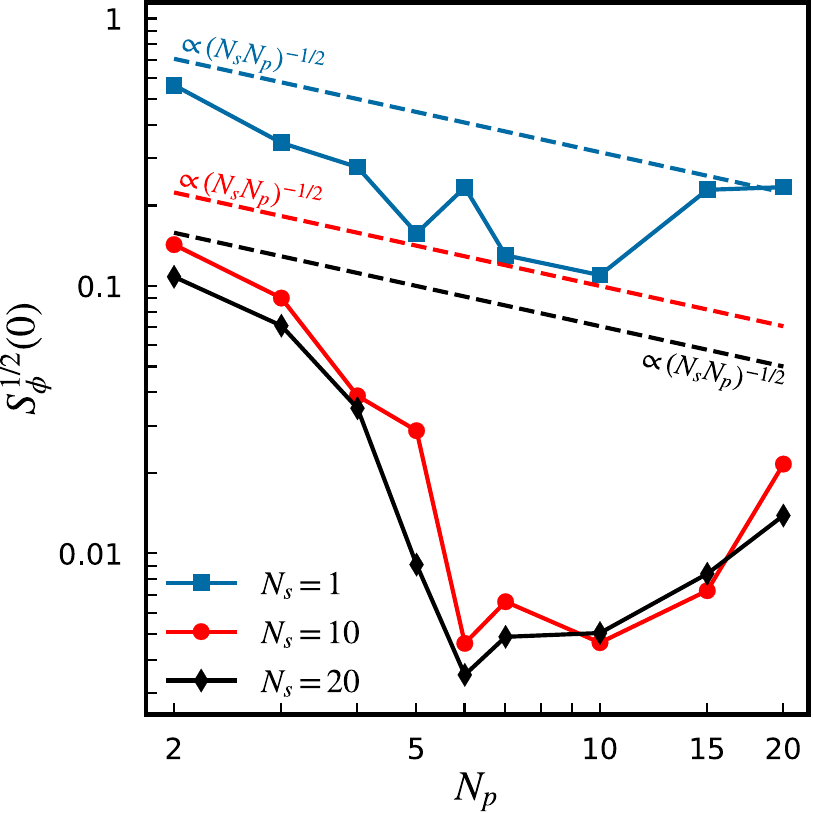}
\caption{Normalized rms flux noise $T=77$ K as a function of $N_s$ and $N_p$ when using the values $(i_b, \phi_a)$ which minimize $S_{\phi}^{1/2}(0)$ instead of maximizing $\overline{v}_{\phi}$.}
\label{fig:Sphi-minimized}
\end{center}
\end{figure}

In the case of SQUID arrays, $S_B^{1/2}$ can be reduced further by increasing $\overline{V}_{B}^{\mathrm{m}\mathrm{a}\mathrm{x}}$. For instance, Fig. \ref{fig:SB-loops} shows $S_B^{1/2}$ as a function of $a$, and we see that increasing both $N_s$ and $N_p$ contributes to a reduction in noise. Furthermore, since $S_V^{1/2}$ is relatively flat with $a$ as shown in Fig. \ref{fig:voltage-loops}, and $\overline{V}_B^{\mathrm{m}\mathrm{a}\mathrm{x}}$ increases with $a$ but plateaus beyond $a=20 \mu$m, $S_B^{1/2}$ in Fig. \ref{fig:SB-loops} stops decreasing for larger $a$. The lowest noise level achieved in Fig. \ref{fig:SB-loops} is $\sim 450 \mathrm{f}\mathrm{T} / \sqrt{\mathrm{H}\mathrm{z}}$ for a (20,20)-array with loop-size $a = 30 \mu$m, which represents a two-fold improvement from the array with loop-size $a = 10 \mu$m. 

It should be noted that variations in the parameters can change the simulation results. From Figures \ref{fig:transfer_function-loops} and \ref{fig:SB-loops}, one sees an increase in the maximum transfer function $\overline{V}_B^{\mathrm{m}\mathrm{a}\mathrm{x}}$ with increasing $\beta_L$ for constant $I_c$, which in turn leads to a decrease in the magnetic field density noise $S_B^{1/2}(0)$. If one instead keeps the loop area constant, but varies the $I_c$ of the JJs; one obtains different results from the ones presented in this paper. For instance, choosing a lower $I_c$ decreases $\overline{V}_B^{\mathrm{m}\mathrm{a}\mathrm{x}}$ but increases the magnetic field density noise $S_B^{1/2}(0)$, while a higher $I_c$ increases $\overline{V}_B^{\mathrm{m}\mathrm{a}\mathrm{x}}$ but lowers $S_B^{1/2}(0)$. This is consistent with the fact that since $\beta_L\sim I_c$ and $\Gamma \sim 1/I_c$, one expects the noise level to decrease as $I_c$ grows larger. This implies one can further improve the device's robustness to noise by increasing the critical current of the junctions.

Lastly, we must discuss the case in which $(i_b, \phi_a)$ are selected to minimize $S_{\phi}^{1/2}(0)$ instead of maximizing $\overline{v}_{\phi}$. Fig. \ref{fig:Sphi-minimized} shows the minimized $S_{\phi}^{1/2}(0)$ as a function of $N_p$ and $N_s$, which exhibits a significantly different trend to Fig. \ref{fig:Sphi-vs-Np}: the rms flux noise has a clear minimum near $N_p\approx 6$ which becomes more prominent for larger $N_s$. At $N_s=20$ and $N_p=6$, $S_{\phi}^{1/2}(0) = 0.0035$ which is over 10 times smaller than in Fig. \ref{fig:Sphi-vs-Np}. However, this comes at the cost of a smaller transfer function: in this case, $\overline{v}_{\phi}/N_s\approx 0.367$ compared to the optimum $\overline{v}_{\phi}/N_s \approx 2.28$ in Fig. \ref{fig:transfer-function}. This shows that the choice of $(i_b, \phi_a)$ can either maximise $\overline{v}_{\phi}$ or minimise $S_{\phi}^{1/2}(0)$, but not both simultaneously.


\section{\label{sec:level1}Conclusion}
In this paper, we have shown through numerical simulations how the noise scales in 1D and 2D SQUID arrays with respect to the number of junctions. In 1D SQUID arrays we observed a $\sim N_s/N_p^{0.3}$ voltage noise spectral density scaling. In contrast, the voltage noise spectral density of 2D arrays follows the $\sim N_s/N_p$ scaling closely. Though increasing $N_p$ beyond a certain value will not further increase the maximum transfer function, it further reduces the voltage noise spectral density. The rms flux noise, which is inversely proportional to the transfer function; deviates from the expected $\sim (N_s N_p)^{-1/2}$ scaling for both the 1D parallel arrays and the 2D arrays when $\bar{v}_{\phi}$ is optimised. Furthermore, we have shown that one cannot optimise the transfer function as well as the flux noise of the array using the same bias current and flux values unless $N_s=1$. By varying the cell size for a given array $(N_s, N_p)$, one can further reduce the magnetic field noise spectral density without compromising the maximum transfer function. This indicates that there is still room for exploration in the improvement of high-$T_c$ SQUID arrays for sensing applications.

\end{document}